\documentstyle[psfig,aps,prl,twocolumn]{revtex}

\begin{document}

\title{The Strange Behavior of Critical Branched Polymers}

\author{H. H. Arag\~ao-R\^ego$^1$, J. E. de Freitas$^{1,2}$, 
Liacir S. Lucena$^1$
\and G. M. Viswanathan$^{1,3}$}

\address{$^1$ International Center for Complex Systems and
Departamento de F\'{\i}sica Te\'{o}rica e Experimental, \\
Universidade
Federal do Rio Grande do Norte, 59078-970 Natal--RN, Brazil}
\address{$^2$Departamento de Matem\'atica, Universidade Federal do Rio
Grande do Norte, 59078-970 Natal--RN, Brazil}
\address{$^3$ Departamento de F\'{\i}sica, Universidade Federal de
Alagoas, 57072-970 Macei\'{o}--AL, Brazil }

\date{\today}

\maketitle
\begin{abstract} We find that 2-dimensional (2-D) critical branched
polymers with no impurities conclusively belong to the same
universality class as 2-D random percolation clusters, although pure
critical 3-D branched polymers do not belong to the 3-D percolation
universality class.  We find, moreover, that the fractal dimension
$d_{\mbox{\scriptsize f}}$ of critical branched polymers is
independent of the presence of a random environment in 2-D, but not in
3-D.  We also report that when there are no impurities the critical
branching probability in 3-D is $b_{\mbox{\scriptsize c}}=3.34 \times
10^{-4} \pm 0.16 \times 10^{-4}$.

\end{abstract}
\bigskip

Polymerization has been an important topic of recent interest in the
physics community\cite{havlin1,prl2,prl3,prl4}. Polymers are
extensively studied~\cite{flory,degennes,doi} complex systems, and can
be either linear or branched~\cite{havlin1,ft,parisi,liacir}.  In
spite of being simpler, linear polymers (LP) present physical features
such as non-Markovian growth and nontrivial scaling. For several
decades it has been known that linear polymers in diluted solutions
can be successfully modeled using self-avoiding walks (SAW).  More
recently it was found that large SAW chains can be more easily
constructed by kinetic growth walks (KGW)~\cite{stanley} that belong
to the same universality class as SAW chains~\cite{peliti}.  
Branched polymers (BP) have a larger degree of complexity and exhibit
an astonishingly rich phenomenology
\cite{havlin1,prl2,prl4,degennes,doi,liacir,impurities,bunde-havlin96,cardy}.
In the 1990's, the KGW approach was generalized to describe the growth
of BP in disordered media\cite{havlin1,liacir,bpgm2}.  Such BP
are typically grown
when polymerization occurs in a solution with two types of
units: (i)~monomers with two chemically active molecular sites
(``tips'') that lead to linear polymerization and (ii)~three-tipped
mononers that lead to branchings (i.e., bifurcations).  
Many important physical properties of the resulting BP
depend on the probability $b$ for branching to occur and the
concentration $c$ of impurities~\cite{impurities} that can block
or slow  the polymerization.

Here, we first address the controversial question: to which
universality class does the critical polymer ($b=b_{\mbox{\scriptsize
c}}$) belong for $c=0$?  A typical phase
diagram~\cite{havlin1,liacir,bpgm2} (Fig.~1(a)) for BP consists of
several distinct regions.  For large $b$
and small $c$,
the polymer is compact and grows indefinitely 
 (i.e., is ``infinite'')
with a
smooth 
(faceted)
growth surface,
while for smaller values $1>b>b_{\mbox{\scriptsize c}}$
closer to the critical line $b_{\mbox{\scriptsize c}}(c)$
 the resulting polymer grows indefinitely but with a rough growth
 surface~\cite{bpgm2}. For yet smaller values of
 $b<b_{\mbox{\scriptsize c}}$ the resulting polymer is finite, due to
 termination of polymerization at all active growth sites.  Along the
 critical line $b=b_{\mbox{\scriptsize c}}(c)$, the polymer is fractal
 (Fig.~1(a)) and has a diverging correlation length.

The commonly held belief that BP belong to the lattice
 animals~\cite{bunde-havlin96} universality class was based on the
 assumption that there is no random environment ($c=0$) and that there
 are only repulsive forces between chains~\cite{cardy}.  It was thus a
 remarkable finding when Bunde {\it et al.}~\cite{havlin1} showed,
 using the branched polymer growth model (BPGM), that critical BP in
 random environments in 2-D belong to the same universality class as
 2-D random percolation clusters.  One conceivable explanation for
 this interesting result is that in 2-D such behavior is induced by
 the random impurity obstacles that, indeed, form real percolation
 clusters with site occupation probability $p=c.$ Hence, it can be
 argued that the polymer grows as if constrained by the random
 obstacles, and this effect can generate an effective attraction
 between chains (Fig.~1(b)).  For the special case $c=0$, however,
 there are no such impurities, hence this physical argument becomes
 irrelevant.  The study by Bunde {\it et al.}~\cite{havlin1} is
 conclusive all along the critical line $b=b_{\mbox{\scriptsize
 c}}(c)$ for $c\neq 0,$ but the important special case of pure BP
 ($c=0$) still remains unconfirmed~\cite{havlin1}, and both the
 percolation as well as the lattice animals universality classes
 remain plausible.
 An additional controversy
 surrounding the special case $c=0$ arises because in many
 systems~\cite{cardy} the introduction of disorder is sufficient to
 modify the universality class (and sometimes even to destroy the
 ordered phase).  The Harris criterion~\cite{cardy} predicts that the
 critical BP in 2-D is a candidate for experiencing a change of
 universality class when going from $c = 0$ to $c \neq 0$.


We investigate this problem by using simulations of the
BPGM~\cite{liacir}.  BPGM is a generalization of KGW~\cite{stanley}
which is able to capture the essential dynamics of branched
polymerization.  We briefly describe the model here (Fig.~1(b)).  The
BP is generated from an initiating seed at time $t=0$
located at the center of a $d$-dimensional lattice. At time step
$t=1$, one of the vacant nearest-neighbor sites of the seed is chosen
randomly and occupied to become the next active growth tip, and the
seed ceases to be an active growth tip.
At each step $t$ of the growth process, a branching also can occur at
every such growth tip, by occupation of an additional (second) site
with probability $b$.  A fraction $c$ of randomly chosen sites is not
available for growth. Hence, at time $t+1$, the polymer can grow from
any of the active tips added at the previous step $t$ to empty
neighboring sites (Fig.~1(b)).

Our extensive 2-D simulations, discussed below, leave no doubt that,
surprisingly, random percolation is indeed the correct universality
class even for $c=0$.  We simulate critical polymers for $c=0$ and
$b=b_{\mbox{\scriptsize c}}(0) \approx 0.056$ and find that the
fractal dimension $d_{\mbox{\scriptsize f}}$, the minimum dimension
$d_{\mbox{\scriptsize min}}$ (Fig.~2(a)) and the chemical dimension
$d_\ell$ (Fig.~2(b)) are in strong agreement with the known values for
a a 2-D critical percolation cluster~\cite{bunde-havlin96}. These
dimensions are defined by the relations $M\sim r^{d_{\mbox{\tiny
f}}}$, $M\sim \ell^{d_\ell}$, and $\ell\sim r^{d_{\mbox{\tiny min}}}$,
where $r$ is radius and $\ell$ is the chemical distance ($\ell=t$ in
BPGM).  Clearly $d_{\mbox{\scriptsize f}}=d_\ell\cdot
d_{\mbox{\scriptsize min}}\approx 1.89$. We conclude that the critical
2-D random percolation cluster and the critical polymer generated by
BPGM unmistakably belong to the same universality class.  Therefore
the fractal dimension of the critical BP in 2-D is independent of the
presence of a random environment---in apparent violation of the Harris
criterion.

We further investigate this ``paradoxical'' result by performing 
similar 3-D simulations of pure ($c=0$) critical BP, because the
Harris criterion favors a little more 
the irrelevance of the quenched
disorder in 3-D.  Until now, such 3-D simulations have not been
possible because the precise value of $b_{\mbox{\scriptsize c}}(0)$ in
3-D was unknown.

So we now determine the value of $b_{\mbox{\scriptsize c}}(0)$ in 3-D
for $c=0$.  The fundamental difficulty of finding a precise value for
$b_{\mbox{\scriptsize c}}(0)$ in 3-D is that the value is extremely
small, $b_{\mbox{\scriptsize c}}(0)<10^{-3}$ (i.e., on average more
than $10^3$ monomer units between successive branchings), making it
computationally expensive to pinpoint it using traditional numerical
approaches that involve tuning $b$.  We overcome this difficulty by
mapping BPGM to the computationally less intensive fixed number of
tips model (FNTM)~\cite{fntm} that has been
shown\cite{fntm,andrade-soc} to generate phase diagrams identical to
BPGM.
FNTM differs from BPGM as follows: rather than fixing the branching
probability $b$, instead FNTM dynamically attempts to fix the number
$N$ of active polymerization growth tips. Hence, the branching process
does not occur with an {\it a priori} fixed value $b$, but rather
branchings occur whenever one or more of the $N$ existing active tips
``die,'' either because of impurities or due to other steric hindrance
effects.  We can be certain that the value of $b_{\mbox{\scriptsize
c}}$ found using FNTM is identical to the one found from BPGM
simulations because FNTM generates polymers in which the number of
active growth tips neither vanishes nor explodes exponentially, i.e.,
corresponding to the critical line at $b=b_{\mbox{\scriptsize c}}$ in
BPGM simulations\cite{fntm,andrade-soc}.  FNTM greatly reduces the
computational burden of generating critical polymers because it
automatically generates critical polymers with $b=b_{\mbox{\scriptsize
c}}(c)$ and, moreover, it has only one free parameter, namely the
impurity concentration $c$. (The arbitrary value of $N>1$ chosen is
irrelevant in the large mass limit, as seen in Fig.~3(a)) The
effective critical branching probability $b_{\mbox{\scriptsize
eff}}=B/M$ is found by dividing the total number $B$ of branchings of
active tips by the total number $M$ of occupied polymer sites.  Most
importantly, FNTM goes spontaneously to the critical line without the
need of parameter tuning (of $b$),
as with BP growth models that exhibit self-organized criticality
(SOC)~\cite{andrade-soc}.

We simulate FNTM for $c=0$ in 3-D and $N= 5,~10,~20,~30,~50,~80$ and
$100$ tips as a function of polymer mass $M$. In the limit $M \rightarrow
\infty$, we find that the values of $b_{\mbox{\scriptsize eff}}=B/M$
converge towards an identical point independently of $N$
(Fig.~3(a)).  The uncertainty for a typical point in
Fig.~3(a) is $\Delta b_{\mbox{\scriptsize eff}}\approx
10^{-5}$. This remarkably small error bar is made possible only
because FNTM dynamically seeks the fixed point attractor near
$b\approx b_{\mbox{\scriptsize c}}$, such that the effective critical
branching probability $b_{\mbox{\scriptsize eff}}=B/M$ fluctuates
around $b_{\mbox{\scriptsize c}}$. By extrapolating
$b_{\mbox{\scriptsize eff}}$ for $1/M
\rightarrow 0$ we find that $b_{\mbox{\scriptsize c}}(0)=3.34
\times 10^{-4} \pm 0.16\times 10^{-4}$.

Since it is not altogether inconceivable that the value of
$b_{\mbox{\scriptsize c}}(0)$ obtained from FNTM will not carry over
to BPGM, we check the above conclusions by simulating BPGM with $2.9
\times 10^{-4} < b < 4.0\times 10^{-4}$. We find (Fig.~3(b)) that the
number of active tips either grows or decays much faster than a power
law except in the approximate range $ 3.3 \times
10^{-4}<b_{\mbox{\scriptsize c}}(0)<3.4 \times 10^{-4}$.  This value
obtained for $b_{\mbox{\scriptsize c}}(0)$ is not inconsistent with
the value $b_{\mbox{\scriptsize c}}=3.34\times 10^{-4}\pm 0.16 \times
10^{-04}$ found using FNTM.

Using this result, we are able to perform 3-D BPGM simulations of pure
critical BP.  We find that $d_{\mbox{\scriptsize f}}\approx
d_{\mbox{\scriptsize min}}\approx 2$ and $d_\ell \approx 1$ in 3-D for
$b=b_{\mbox{\scriptsize c}}(0)$ and $c=0$ (Figs.~2(c),~(d)),
definitely ruling out the 3-D random percolation universality class.
Although this is the expected field theoretical result (see
ref.~\cite{cardy}) for 3-D polymers, the Harris criterion indicates a
relevant disorder and a change in the value of $d_{f}$ (because $d \nu
= 1.5 < 2$ for $d=3$ BP).  This prediction is
confirmed by the finding of Bunde {\it et al.}~\cite{impurities}, that
$d_{\mbox{\scriptsize f}} \approx 2.53$ for $b_{\mbox{\scriptsize
c}}=1$, consistent with 3-D random percolation.

Somewhat surprising is our confirmation that in 2-D the fractal
dimensions of the critical BP are identical for the pure case $c=0$
and the maximally disordered case $b=1$, $c\approx
1-p_{\mbox{\scriptsize c}}$ (see Fig.~1(a)), where
$p_{\mbox{\scriptsize c}}\approx 0.59 $ is the critical site
occupation probability in random percolation for a square lattice.
For $b=1$ the critical polymer grows in a vacant space that is itself
almost a critical percolation cluster (Fig.~1(a)).  Since there is no
random environment for $c=0$, a larger fractal dimension had been
conceivable in principle.  Our findings indicate, strangely, that the
fractal dimension (but not the lacunarity) of the critical polymer in
2-D is independent of the growth environment. Why this is so is indeed
an interesting question that merits further investigation, and seems
to be related to a self-organizing process~\cite{andrade-soc}.
Criticality occurs only when there is a delicate balance between the
rates of tip deaths and branchings---a dynamic not unlike the one used
in the self-organized generation of infinite 2-D percolation
clusters~\cite{adriano}.

We now comment briefly on our finding that, strangely, pure critical
BP fall into the percolation universality class in 2-D but not in 3-D.
The deeper reason behind this mysterious behavior is possibly related
to the well known fact that in 2-D, encounters between different
branches of a BP can result either in ``trapping'' of a polymer chain
or else in ``scattering'' (bending) of the chains. In 3-D a completely
different effect becomes important in the large-scale growth dynamics:
interpenetration and ``entanglement'' of BP
chains~\cite{degennes}---not by independent linear chains, but by
linear parts of a BP that act independently at criticality, because
the chemical distance between successive branchings is much larger
than the persistence length.  In 2-D, there is no entanglement (for
topological reasons), but in 3-D such effects can counteract the
excluded volume repulsion, so that the mean square displacement of the
tips grows linearly with mass (as with polymer melts, see Ch.~2 and~8
of ref.~\cite{degennes}), leading to power law scaling equivalent to
ideal Gaussian chains with $d_{\mbox{\scriptsize f}}=2$.  Effectively,
there is a larger repulsion between chains in 2-D, thereby leading to
superdiffusive motion of the tips (with $d_{\mbox{\scriptsize f}}<2)$.
Note that trapping in 3-D is extremely rare, hence the very low value
$b_{\mbox{\scriptsize c}}(0)$. 

Finally, one hint for the cause of the observed change in the
universality class (when we vary $b$ from $b=b_{\mbox{\scriptsize
c}}(0)$ to $b=1$) in 3-D becomes evident by comparing the two
situations with zero and maximum disorder: for $b_{\mbox{\scriptsize
c}}(0) = 3.3 \times 10^{-4} $ the ``freedom'' of motion of the tips is
limited only by the sparse steric hindrance due to the polymer chains, 
while for $b=1$ the
behavior is dictated by the intersticial percolation cluster available
for BP growth.  The arguments presented above raise new questions,
such as whether other critical exponents vary along the critical line
$b=b_{\mbox{\scriptsize c}}(c)$.

We thank U. L. Fulco, H. J. Hilhorst, M.~L.~Lyra, S. Roux, and
L. R. da Silva for discussions; CNPq, PRONEX, the Projeto Nordeste de
Pesquisa \& FINEP for support.

\begin{figure}
\label{intro}
{
\hspace{0.1in}
\psfig{figure=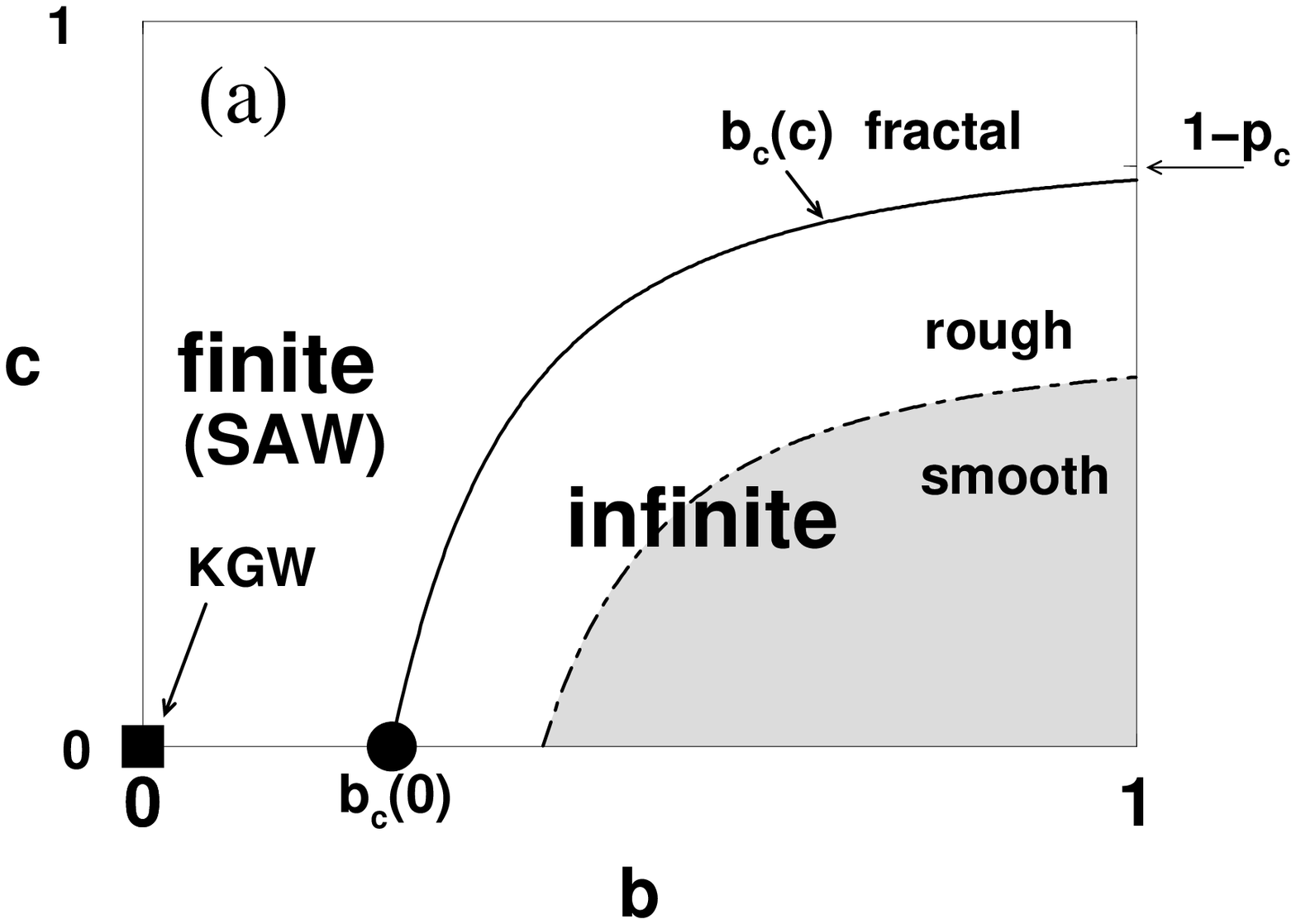,height=1.8truein}

~

\vspace{0.3in}
~~~~~\psfig{figure=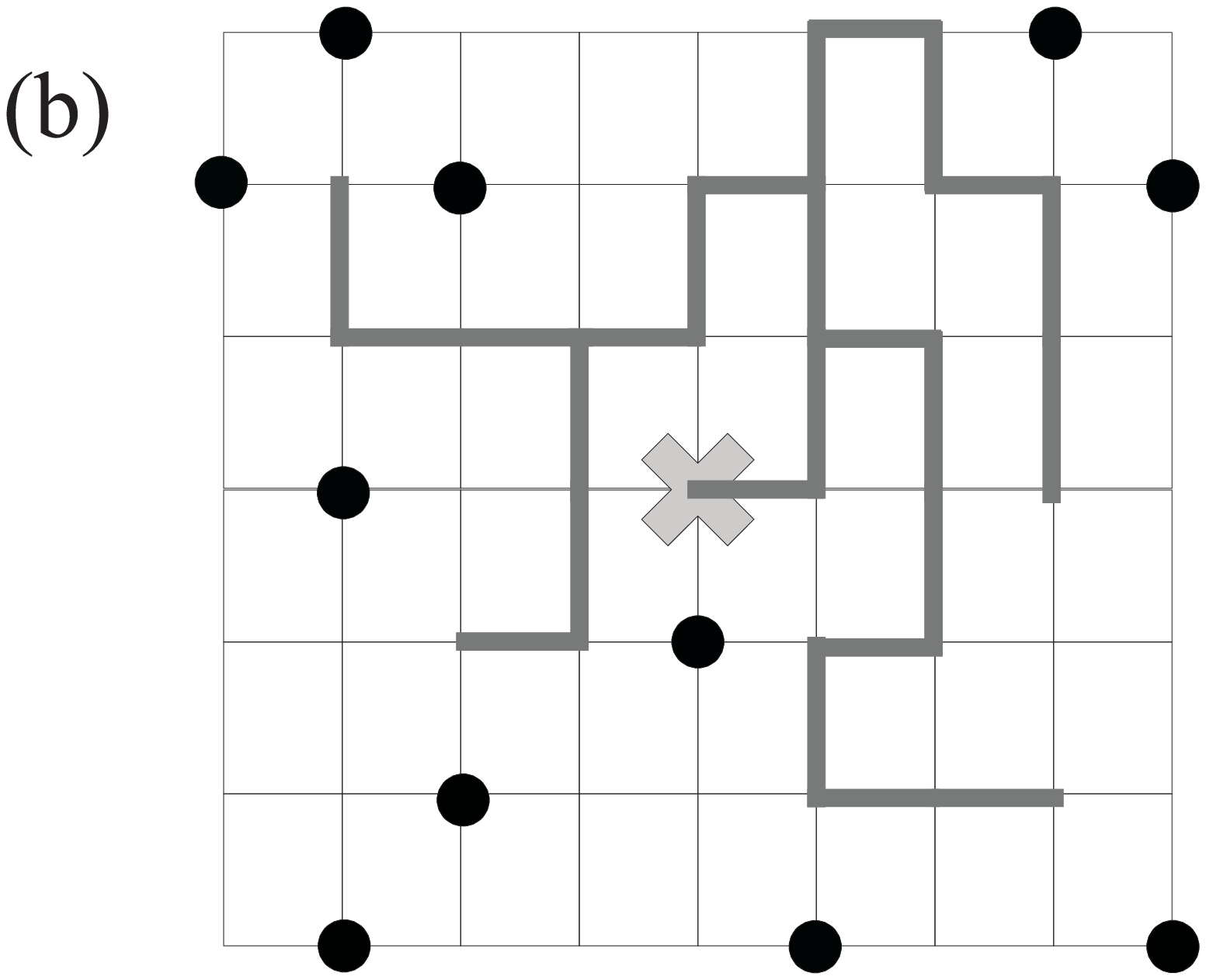,height=1.8truein}}
\caption{ (a)~schematic phase diagram for BP (see text).  We focus
exclusively on the critical point at $b=b_{\mbox{\scriptsize c}}(0)$
that corresponds to BP growth in the realistic case of no impurities.
(b)~the branched polymer growth model (BPGM) assumes that growth takes
place only at active growth tip sites located on a lattice. The seed
(cross) initiates polymerization governed by a branching probability
$b$.  When the impurity concentration $c$ is nonzero, then the random
obstacles (circles) can sometimes cause an effective attraction
between chains (see branches on the right), but this effect disappears
when $c=0$.}

\end{figure}

\begin{figure}
\label{percolation}
\centerline{~~~~\psfig{figure=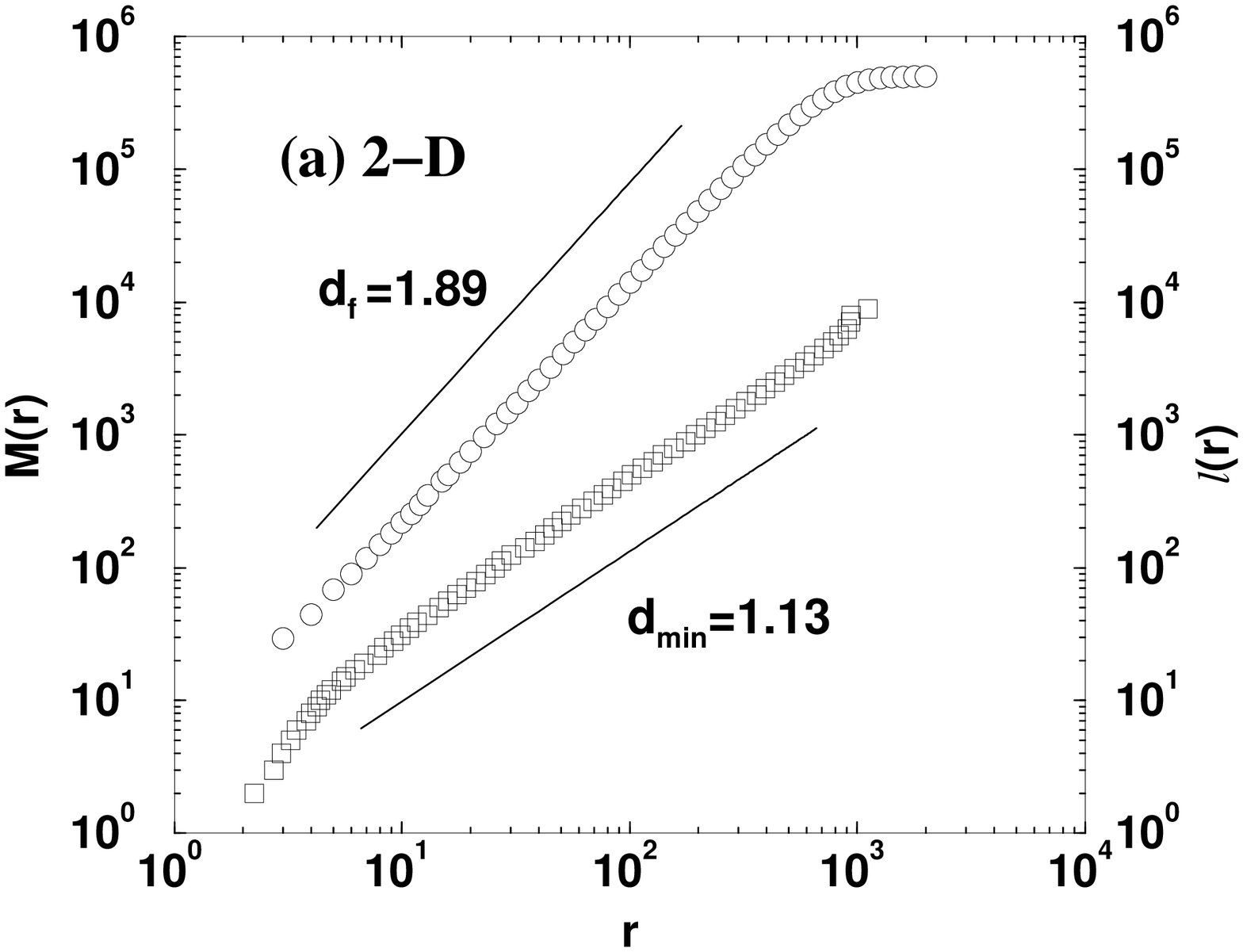,height=2.2truein}}
\centerline{\psfig{figure=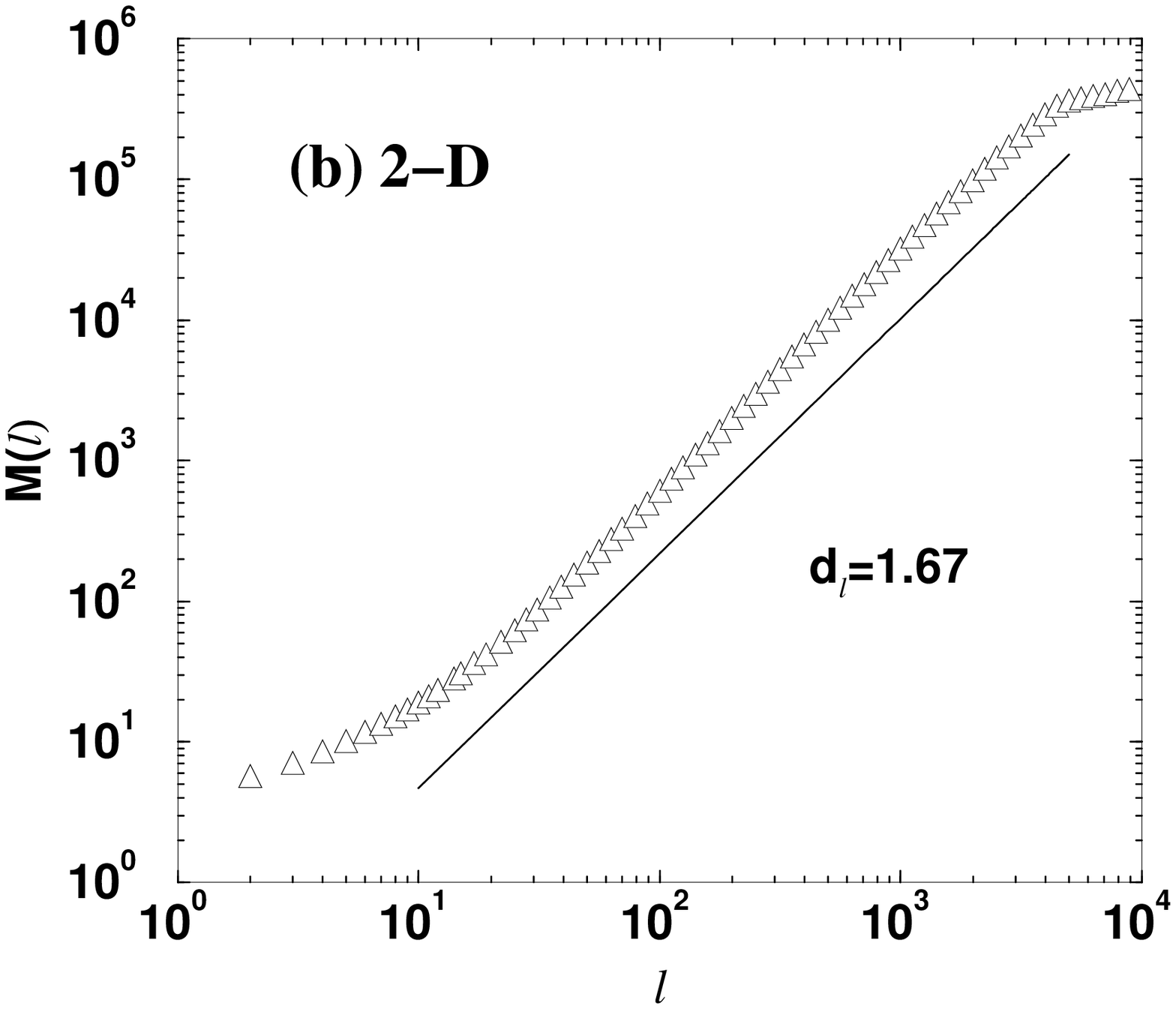,height=2.1truein}}
\centerline{~~~~\psfig{figure=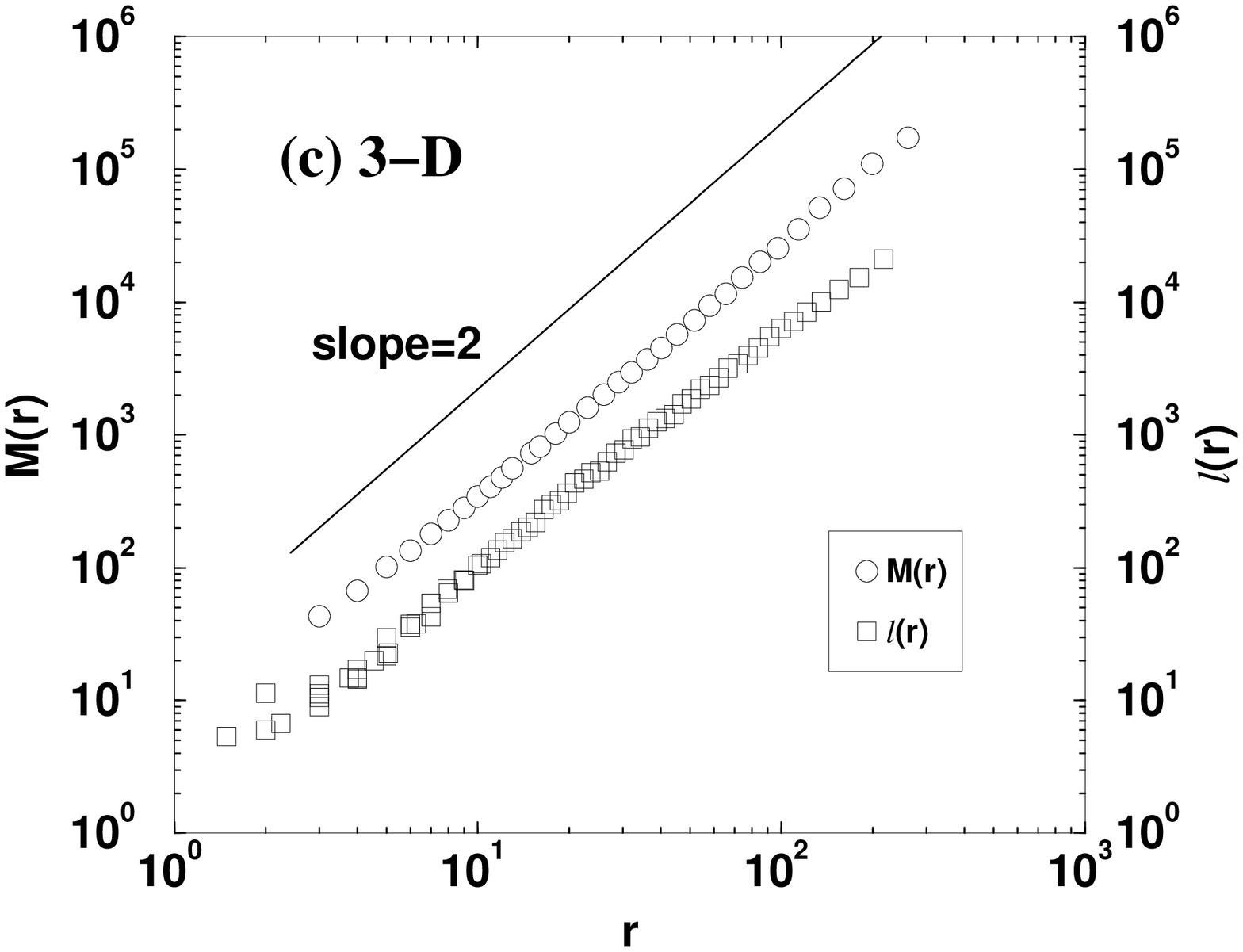,height=2.2truein}}
\centerline{\psfig{figure=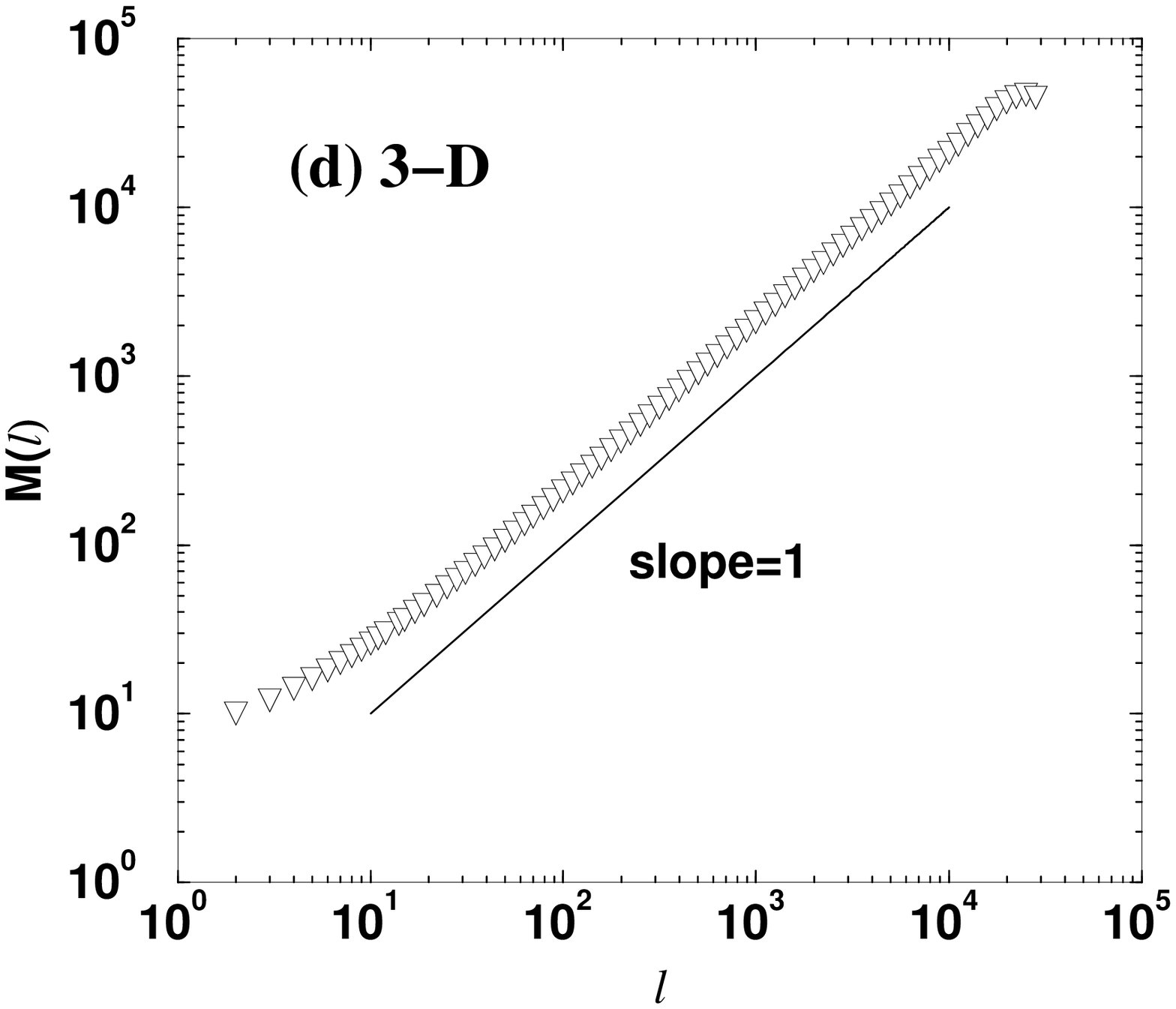,height=2.1truein}}
\caption{(a)~double-log plot of the polymer mass in units of monomers
(mass axis label at left) and chemical distance (label at right) as a
function of radius for BP with critical branching probability
$b_{\mbox{\scriptsize c}}(0)$ in the absence of impurities in 2-D. We
have used 100 samples with mass $M=5\times 10^5$ monomers. The fractal
dimension $d_{\mbox{\scriptsize f}}$ and the minimum dimension
$d_{\mbox{\scriptsize min}}$ (the slopes) are those of a 2-D critical
percolation cluster (straight lines). (b) double-log plot of the
polymer mass as a function of the chemical distance for the identical
simulations, showing that the chemical dimension is also consistent
with the universality class of 2-D random percolation.  (c)~polymer
mass and chemical distance vs. radius and (d)~polymer mass vs. 
chemical
distance in 3-D, showing that the behavior is definitely different
from 3-D random percolation ($d_{\mbox{\scriptsize f}}=2.52$,
$d_\ell=1.84$, $d_{\mbox{\scriptsize
min}}=1.37$~\protect\cite{bunde-havlin96}).  We have averaged over 900
samples.  We find that 3-D critical BP scaling is more reminiscent of
polymer melts~\protect\cite{degennes} than random percolation.}

\end{figure}

\begin{figure}
\label{3d}
\centerline{~~~\psfig{figure=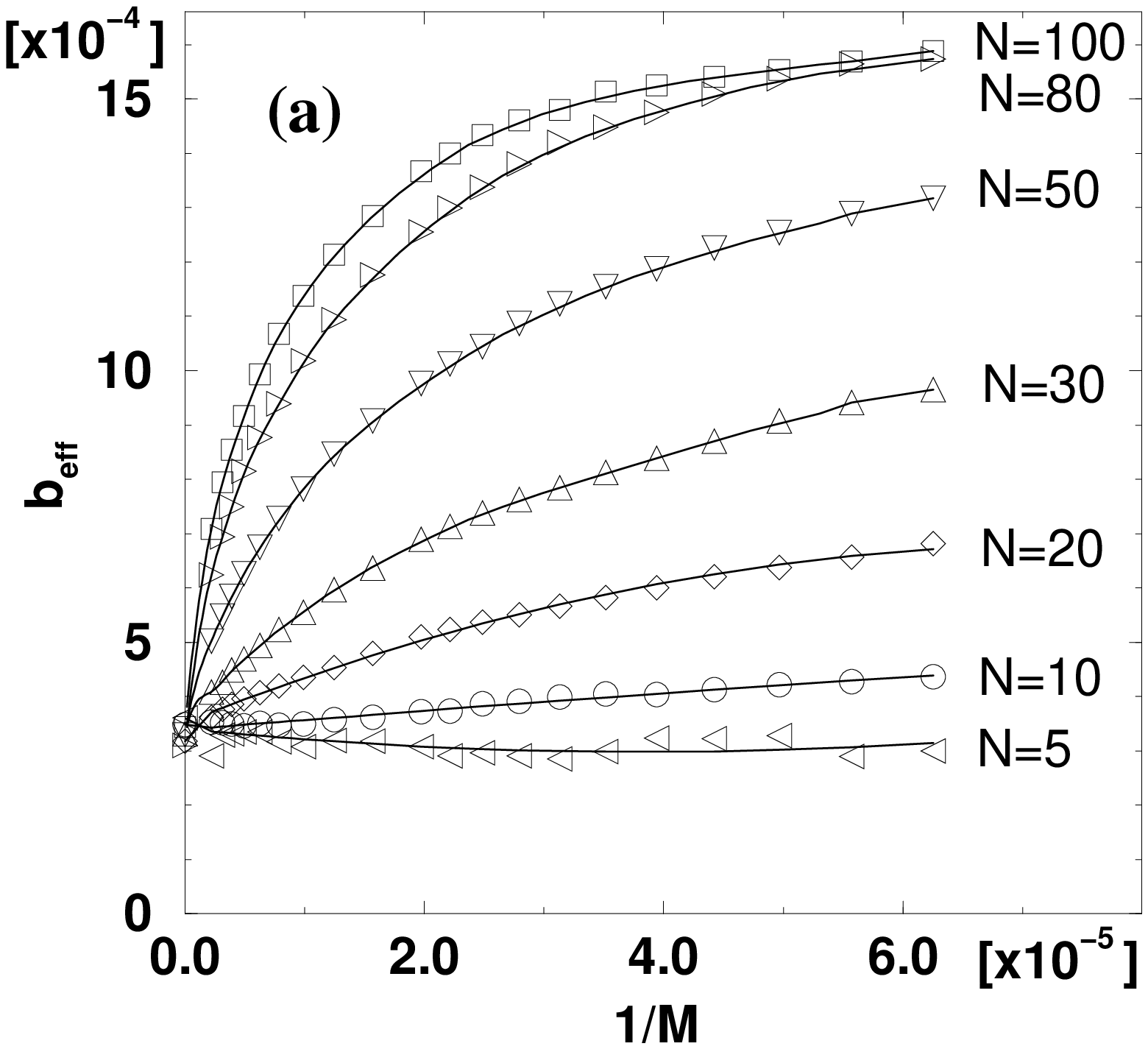,height=2.2truein}}
\centerline{\psfig{figure=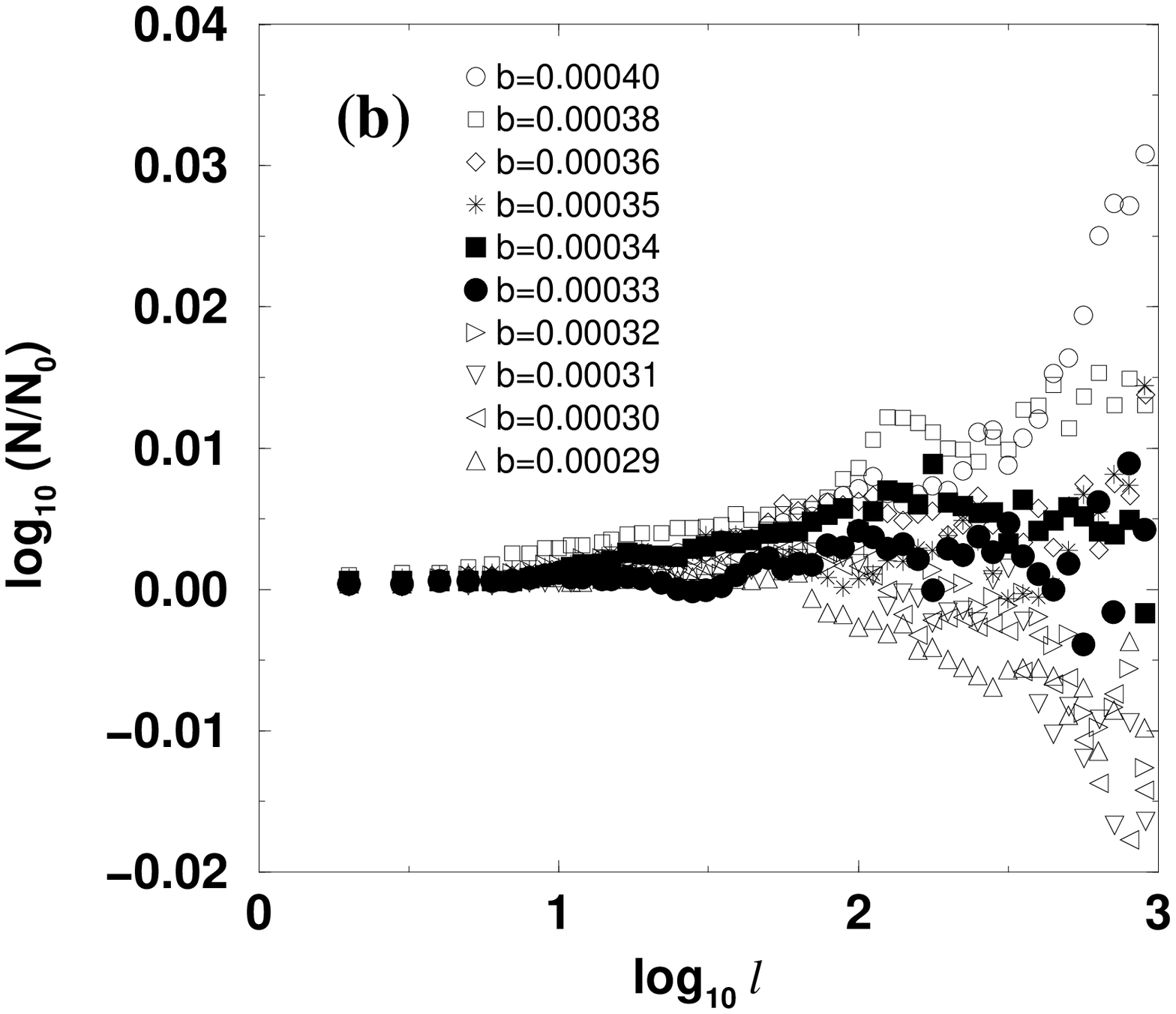,height=2.2truein}~}
\caption{(a) $b_{\mbox{\scriptsize eff}}$ as a function of $1/M$ and
the number $N$ of active growth tips for FNTM in 3-D.  The solid lines
are fitted curves, by which we are able to extrapolate for arbitrarily
large masses. The values shown for $1/M=0$ are the extrapolated
results. We are thus able to estimate the value of
$b_{\mbox{\scriptsize c}}(0)=3.34 \times 10^{-4} \pm 0.16 \times
10^{-4}$.  (b)~double log plot of $N$ as a function of chemical
distance $\ell$ for BPGM, with $N_0=2$ tips initially at $t=0.$ For
values $b$ larger than the estimated value of $b_{\mbox{\scriptsize
c}}(0)$, we find that $N$ tends to explode exponentially, while for
$b<b_{\mbox{\scriptsize c}}(0)$ we find $N$ decays. For $ 3.3 \times
10^{-4}< b < 3.4 \times 10^{-4}$, however, we obtain power law growth
for $N$, indicating critical branching.}

\end{figure}

\end{document}